\documentclass[sigconf,natbib=true,anonymous=false,review=false]{acmart}
\AtBeginDocument{%
  }
\usepackage{latexsym}
\usepackage{graphicx}
\graphicspath{{./images/}}
\usepackage{booktabs} 
\usepackage{color}  
\usepackage{amsmath}  
\usepackage{subcaption}
\usepackage{caption}
\usepackage{tikz}
\usepackage{pifont}
\usepackage{colortbl} 
\usepackage{framed}
\usepackage{multirow}
\usepackage{multicol}
\usepackage{hyperref}
\usepackage{url}
\usepackage{balance}
\usepackage{verbatim}
\usepackage{cancel}
\usepackage{xspace} 
\usepackage[ruled,vlined]{algorithm2e}
\usepackage{bbold} 
\usepackage{comment}
\usepackage{CJKutf8} 

\newcommand{\phrase}[1]{\textit{``#1''}}

\definecolor{urlcolor}{HTML}{0645AD}

\newcommand{\squishlist}{
    \begin{list}{$\bullet$}{ 
        \setlength{\itemsep}{0pt}
        \setlength{\parsep}{1pt}
        \setlength{\topsep}{1pt}
        \setlength{\partopsep}{0pt}
        \setlength{\leftmargin}{1.5em}
        \setlength{\labelwidth}{1em}
        \setlength{\labelsep}{0.5em} 
    } 
}

\newcommand{\squishend}{
  \end{list}  }

\newcommand{\benchmark}{\textsc{DrugRec}\xspace}
\newcommand{\explainmr}{\textsc{TraceDR}\xspace}
\newcommand{\method}{\textsc{TraceDR}\xspace}

\newcommand{\vspaceopt}{\vspace*{0.1cm}}

\copyrightyear{2025} 
\acmYear{2025} 
\acmConference[CIKM '25]{the 34nd ACM International Conference on Information and Knowledge Management}{November 10--14, 2025}{Seoul, Korea}

\begin{document}

\title{Traceable Drug Recommendation over~Medical~Knowledge~Graphs}


\author{Yu Lin}
\affiliation{%
  \institution{Southwest Jiaotong University}
  \city{Chengdu}
  \country{China}}

\author{Zhen Jia}
\affiliation{%
  \institution{Southwest Jiaotong University}
  \city{Chengdu}
  \country{China}}

\author{Philipp Christmann}
\affiliation{%
  \institution{Max Planck Institute for Informatics
  }
  \city{Saarbruecken}
 \country{Germany}}

\author{Xu Zhang}
\affiliation{%
  \institution{Southwest Jiaotong University}
  \city{Chengdu}
  \country{China}}

\author{Shengdong Du}
\affiliation{%
  \institution{Southwest Jiaotong University}
  \city{Chengdu}
  \country{China}}

\author{Tianrui Li}
\affiliation{%
  \institution{Southwest Jiaotong University}
  \city{Chengdu}
  \country{China}}

\renewcommand{\shortauthors}{Yu Lin et al.}
\newcommand{\myparagraph}[1]{\noindent \textbf{#1}.}
\setcounter{secnumdepth}{4}

\begin{abstract}
Drug recommendation (DR) systems aim to support healthcare professionals in selecting appropriate
medications based on patients' medical conditions.
State-of-the-art approaches utilize deep learning techniques for improving DR,
but fall short in providing any insights on the derivation process of recommendations
-- a critical limitation in such high-stake applications.
We propose \explainmr, a
novel
DR system operating over a medical knowledge graph (MKG),
which ensures access to large-scale and high-quality information.
\explainmr simultaneously predicts drug recommendations
and related evidence within a multi-task learning framework,
enabling \textit{traceability} of medication recommendations.
For covering a more diverse set of diseases and drugs than existing works,
we devise a framework for automatically constructing patient health records 
and release \benchmark, a new large-scale testbed for DR.

\end{abstract}

\maketitle

    	
    	        

\begin{figure*} [t]
     \includegraphics[width=\textwidth]
     {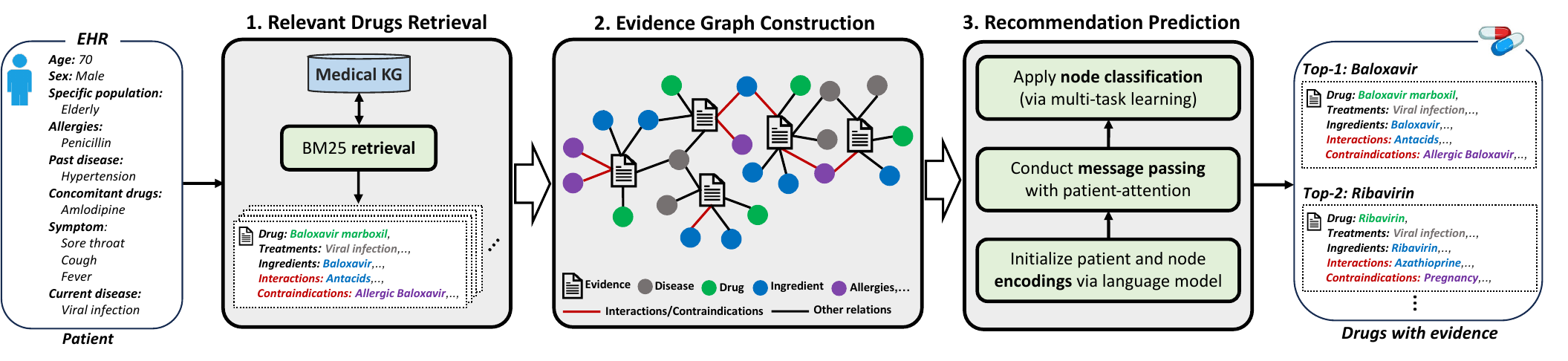}
     \vspace*{-0.9cm}
     \caption{
     Overview of \explainmr's three stages:
     \explainmr first retrieves candidate drugs via BM25,
     for which an evidence graph with relevant medical information is constructed.
     Our GNN, with dedicated patient-attention, scores drugs in the graph.
     }
     \label{fig:overview-explaindr}
     \vspace*{-0.2cm}
\end{figure*} 

\vspace*{-0.3cm}
\section{Introduction}
\label{sec:intro}
\myparagraph{Motivation}
Automatic drug recommendation (DR) is an essential task in healthcare,
assisting doctors with drug suggestions and improving prescribing efficiency. 
In addition, DR helps the public identify suitable over-the-counter medications based on their symptoms, reducing the risks of self-medication.


There has been substantial research on automated drug recommendations~\cite{ali2023deep}.
{Most methods~\cite{choi2016retain, gong2021smr, shang2019gamenet,kim2024vita, yang2021safedrug, yang2023molerec, chen2023context}
leverage recent advances in deep learning and train prediction models from patients' electronic health records (EHR).
These approaches rely heavily on historical records, which makes them less geared for rare diseases or single-visit patients~\cite{zhao2024leave}.} Further, their inherent reliance 
on \textit{black-box} neural methods makes it difficult for users to understand the rationale behind recommendations:
it remains unclear, based on which information a certain drug was predicted as relevant.
Such \textit{traceability} is of utmost importance in any health-related application ~\cite{fan2022comprehensive}.
Recommendations for specific medications should be supported by appropriate and high-quality \textit{evidence},
enabling practitioners or even end users (if must be) to investigate the correctness of drug recommendations.

\vspace*{0.05cm}
\myparagraph{Approach}
To overcome these limitations, 
we propose \explainmr
(\underline{Trace}able\,\underline{D}rug\,\underline{R}ecommendation\,over\,Medical\,Knowledge\,Graphs):
%
%
%

\vspace*{0.1cm}

\noindent (i) Given a 
patient's {current} disease and 
medical conditions,
we 
identify a set of candidate \textit{drugs},
and retrieve related 
data
from a medical KG, {alleviating the data sparsity of rare diseases.}

\noindent (ii) 
We construct an \textit{evidence graph}, with \textit{drugs}, 
related \textit{diseases}, \textit{ingredients} and \textit{contraindications} as nodes,
as well as 
evidence nodes representing 
information
on the drugs.
Evidence nodes are connected to 
{drugs}, {diseases}, {ingredients}, and {contraindications}.

\noindent  (iii) 
We devise a graph neural network (GNN) based on a novel \textit{patient-attention} mechanism,
to 
jointly
predict drug recommendations and the corresponding evidence 
in
a multi-task learning framework.
Such 
evidence serves as an \textbf{explanation} for the 
recommendation,
capturing the derivation of the prediction.



\vspace*{0.05cm}
\myparagraph{A new large-scale benchmark}
Existing DR datasets~\cite{johnson2016mimic,johnson2020mimic,he2022dialmed} fall short along one or multiple of the following dimensions:\\
\noindent (i) little diversity of diseases and patients,\\
\noindent (ii) data is of a small-scale (in the order of a few hundred), or\\
\noindent (iii) lack of medical context (e.g., no patient information).\\
To address these shortcomings, we devise a framework to construct DR data at a large scale,
via an automated 
construction process.
The framework
leverages an MKG 
and a large language model (LLM) to generate data automatically.
As a result, we release \benchmark, comprising 
$21$k patient data,
and a diverse set of $14$k diseases.

Experiments on \benchmark demonstrate substantial performance benefits
of \explainmr across different metrics,
compared to a wide family of baselines:
from LLMs
to state-of-the-art DR methods,
standard retrieval techniques,
and
GNNs.
Notably, drug recommendations by
\explainmr are \textit{traceable}
to the relevant medical information.

\vspace*{0.1cm}
\myparagraph{Contributions}
The salient contributions of this work are:\footnote{All code and data are available at \textcolor{urlcolor}{ \textbf{\url{https://github.com/zhenjia2017/TraceDR}}}.}
\squishlist
\item proposing \explainmr, the first DR system using a large-scale MKG,
enabling access to vast amounts of high-quality data;

\item devising a methodology that provides traceable drug recommendations by design;



\item releasing \benchmark, a large-scale benchmark covering a wide range of diseases and drugs.
\squishend
\newpage

\section{The \explainmr Methodology}
\label{sec:method}
\begin{CJK*}{UTF8}{gbsn}
Figure~\ref{fig:overview-explaindr} provides an overview of \explainmr,  which has three main stages: (i) relevant drugs retrieval, (ii) evidence graph construction, and (iii) recommendation prediction.






\vspace*{-0.2cm}
\subsection{Relevant Drugs Retrieval}
\label{subsec:retrieval}
The first stage retrieves drug candidates from an MKG~\cite{liao2022medical},
and derives related evidence.
We use
the standard IR model
BM25~\cite{robertson2009probabilistic} for retrieving drug candidates (with $k1$=$1.5$ and $b$=$0.75$).
Each drug is represented by the corresponding target diseases, ingredients and contraindications,
as captured in the MKG.
The patient's disease and symptoms 
are utilized as query,
and the top-$50$ drugs are kept.
We further add
concomitant drugs (i.e., medications currently being taken), 
as provided in the patient's health record,
for context.
Information
about such drugs
can be a key indicator in the recommendation process,
to avoid 
adverse
drug-drug interactions (DDIs).
Finally, we utilize the MKG to retrieve key information,
like
\textit{treatments}, \textit{ingredients} or \textit{interactions},
from the 1-hop neighborhood
of all candidate drugs
(top-$50$ BM25 results and concomitant drugs).
For every drug, this information is
verbalized~\cite{christmann2023explainable}
into text,
and serves as evidence for supporting recommendations
in subsequent stages of our methodology.


\vspace*{-0.2cm}
\subsection{Evidence Graph Construction}
\label{subsec:graph}
We construct an \textit{evidence graph} with the following nodes:\\
\noindent (i) \textit{entity} nodes, including \textit{drugs}, \textit{diseases}, \textit{drug ingredients}
or \textit{contraindications}, that are represented by the corresponding label, and\\
\noindent (ii) \textit{evidence} nodes, which contain the verbalized MKG facts for each drug, representing
the drugs' essential information.\\
An evidence node is connected with all the entity nodes
that are mentioned in the verbalized text.
For example, if an evidence node mentions a certain drug ingredient,
a connection to this ingredient is added.
Figure~\ref{fig:overview-explaindr} shows an example of the derived graph structure.

\vspace*{-0.2cm}
\subsection{Recommendation Prediction}
There has been plenty of work on
GNNs
in the context of recommender systems and IR in general~\cite{gao2023survey, ying2018graph, christmann2023explainable, chen2023heterogeneous}.

Most of these works operate solely on the graph itself,
without considering additional information (e.g., textual data).
In our setting, however, it is essential to utilize information 
about the patient for making informed decisions.
To this end, we devise
a novel \textit{patient-attention} mechanism, that is applied within
each message passing step of the GNN.
This
mechanism
computes the attention of graph nodes to
the patient EHR which is not part of the graph.
The patient-attention thus
captures the relevance of graph nodes to the
patients and their medical conditions
within each step of the recommendation derivation.


\vspaceopt
\myparagraph{Patient and node encodings}
We represent a patient by concatenating
their
age, sex, 
specific population (e.g., \phrase{pregnant}), allergies, current disease, symptoms,
past diseases and concomitant drugs, utilizing a separator ("$\|$").
Graph nodes are represented by their textual surface forms.
We then use a
language model (LM)\footnote{\textcolor{urlcolor}{\url{https://huggingface.co/hfl/chinese-roberta-wwm-ext}}}
to obtain the initial encodings
(with a dimension of $768$)
of
patients and graph nodes,
that are
the inputs to our GNN.

\vspaceopt
\myparagraph{Patient-attention}
The core part of a GNN is the \textit{message passing},
which distributes information among neighboring nodes in the graph.
Within the message passing procedure,
we integrate our new patient-attention mechanism to re-weight messages
based on their relevance to the patient data.
This relevance is measured using the dot product
of the patient encoding and the respective node encoding.
The key idea is that information
%
that is
more relevant to the patient,
is prioritized in the encoding updates.

We compute patient-attention ${\alpha_{p}}$ for a node $n$ as:
\begin{equation}
{\alpha_{p}}=\underset{n' \in \mathcal{N}(n)}{softmax}(W_a(\vec{n})\cdot \vec{p})
\end{equation}

where $\vec{p}$ and $\vec{n}$ are patient and node encodings, respectively.
Node encodings are linearly projected via $W_a$, and then multiplied by $\vec{p}$
to derive a scalar.
The softmax function is then applied among all neighboring nodes ($n' \in \mathcal{N}(n)$),
to form a probability distribution.

\vspaceopt
\myparagraph{Updating node encodings}
The encodings of graph nodes are updated based on the weighted messages from neighboring nodes:
\begin{equation}
    \vec{n} = \bigg( \vec{n} + W_m  (\sum\nolimits_{n' \in \mathcal{N}(n)} \alpha_{p} \cdot \vec{n'} ) \bigg)
\end{equation}
where $W_m$ is used for linear projection of messages.
The update procedure is the same for entity and evidence nodes in the graph.




\vspaceopt
\myparagraph{Multi-task learning}
The GNN
unifies
two node-classification tasks,
jointly predicting scores for drugs and supporting evidence~\cite{christmann2023explainable}.
After applying message passing,
we utilize bi-linear layers operating on the updated entity/evidence and patient encodings.
Scores are transformed
into a probability distribution
via a softmax function,
for each of the two node types.
For both tasks, we employ binary-cross-entropy
over the predicted scores
as the loss function.
The final loss for training the GNN is a weighted sum
to balance the multi-task learning objective. 


\end{CJK*}
\label{subsec:prediction}


\vspace*{-0.3cm}
\section{The \benchmark Benchmark}
\label{sec:benchmark}
\vspace*{-0.1cm}
\subsection{Limitations of Existing Benchmarks}
Popular datasets for DR, specifically
\textsc{MIMIC-III}~\cite{johnson2016mimic}
and \textsc{MIMIC-IV}~\cite{johnson2020mimic},
collected EHRs in the intensive care unit of a hospital,
limiting the data to highly critical diseases in specific contexts.
In addition, data about symptoms is absent,
which can provide critical insights into diseases~\cite{tan20224sdrug}.
The DIALMED benchmark \cite{he2022dialmed} 
collected and annotated dialogues between doctors and patients from a medical consultation website.
Dialogues are primarily related to patients' current conditions, excluding demographics, past diseases, 
and the medications they are currently taking, the crucial information for DR to minimize the risk of adverse drug reactions and avoid harmful DDIs.
Further, DIALMED has small-scale diagnoses and medications, unable to capture a wider variety of diseases.  


\vspace*{-0.2cm}
\subsection{Data Construction Framework}

\vspace*{-0.1cm}
To overcome these limitations, we devise a framework to construct a new dataset \benchmark. Inspired by~\cite{fansi2022ddxplus} (for the different task of diagnosis), we propose a method to automatically construct synthetic EHRs.
An EHR contains
(i) patient information (sex, age, allergies, etc.),
(ii) past diseases and concomitant drugs,
(iii) current disease, symptoms, and recommended drugs.



\vspaceopt
\myparagraph{Patient information}
{For each synthetic \textit{patient}, the age is generated from the population age distribution based on real census data~\cite{walonoski2018synthea}.}
We then set the sex of the patient under the 1:1 ratio of \phrase{male} to \phrase{female}.
In our benchmark, 
we also add relevant medical conditions for patients
that are important to consider for DR:
 pregnant women ($3.5\%$), breastfeeding women ($5.4\%$), and populations with reduced liver (${2.9}\%$) or renal function (${9.4}\%$). 
Finally, for $20\%$ of patients, we specified
allergies to specific drugs or ingredients, from potential allergens in the MKG.

\vspaceopt
\myparagraph{Current disease and symptoms}
We randomly assign a disease to each patient from the MKG, ensuring that no disease is assigned to more than two patients.
{We start with the simple one-disease scenario to verify the effectiveness of our method, note that it is straightforward to extend our work to a multi-disease situation.}
The diseases not in line with the patient's demographics (e.g., due to age or sex) should be disregarded.
{For example, \phrase{Infantile Paralysis} is a disease that only children suffer from, and \phrase{Prostatitis}
only affects males.
} 
{We include an LLM, Qwen~\cite{yang2024qwen2}\footnote{\textcolor{urlcolor}{\url{https://help.aliyun.com/zh/model-studio/getting-started/models}, using \texttt{Qwen-Max}}} with in-context learning (ICL)~\cite{brown2020language} in the pipeline to filter inapplicable diseases,
which operates on top of a rule-based filter to improve data quality.
We provide $8$ demonstrations containing patients' information, with diseases, incorrect (or correct) labels.} 



To generate symptoms for the disease, we again utilize the Qwen LLM with ICL. We hand-crafted $8$ demonstrations containing basic patient information, disease, and symptoms.
For example, \phrase{Cough, phlegm, fever} would be  the generated symptoms for a 35-year-old male patient with \phrase{respiratory tract infection}.

\vspaceopt
\myparagraph{Past diseases, concomitant drugs, and recommended drugs}
We first construct a candidate recommendation set for treating the current disease. 
Then we randomly sample $1$-$3$ drugs violating DDIs with at least one of the candidate recommendations as concomitant drugs. 
This poses a challenge for DR methods, which have to carefully consider any such cases to avoid conflicts with DDIs.
 
The diseases connected to these concomitant drugs are added to the patient's list of past diseases. {To ensure that diseases are in line with patients' demographics, we disregard inapplicable diseases using the same method for current diseases.}


Finally, we prune out candidate drugs that have DDIs
with the concomitant drugs from our set of recommendations. 
We also prune both concomitant and candidate drugs via safe drug use rules~\cite{liao2022medical}
to ensure 
realistic scenarios, leaving the remaining concomitant drugs and candidates as the patient's medical history and ground-truth recommendations, respectively.

\vspaceopt
\myparagraph{Benchmark characteristics} We construct \benchmark with {$21{,}000$} EHRs,
which are split in a $60:20:20$ ratio into the
train ($12{,}600$),
dev ($4{,}200$) and test set ($4{,}200$).
The benchmark covers $100$k drugs (appear $15.10$ times on avg.) and $14$k diseases (appear $1.54$ times on avg.),
with maximum $2$ patients suffering from the same disease.
All data is in Chinese.
Table~\ref{tab:statistics} shows the basic statistics. 

\vspaceopt
\myparagraph{Quality analysis} 
We randomly sampled $300$ EHRs from the benchmark and manually evaluated their correctness. Notably, we found that {$94.4\%$} of
of the EHRs are fully correct and realistic.
We identified the following errors:
{(i) incorrect patient information ($1.3\%$),}
(ii) incorrect diagnosis {($2\%$)},
or (iii) incorrect medical history {($2.3\%$).}

\begin{table} [t]
    \footnotesize
	\centering
	\caption{Basic statistics for the new \benchmark benchmark.}
    \vspace*{-0.4cm}
	\resizebox*{0.9\columnwidth}{!}{
         \begin{tabular}{p{3.5cm} p{3.6cm}}
            \toprule
                \textbf{Patients / EHRs} & $21{,}000$ ($10{,}358$ female, $10{,}642$ male) \\
                \textbf{Diseases} & $14{,}023$ \\
                \textbf{Drugs} & 
                $100{,}186$  \\
            \midrule
                \textbf{Children}        & $960$ ($<12$ years old)\\
                \textbf{Elderlies}         & $3{,}775$  ($\geq65$ years old)  \\
            \midrule
                \textbf{Allergies} & $4{,}113$ \\
                \textbf{Reduced liver or renal function} & $2{,}589$ \\
                \textbf{Breastfeeding} & $1{,}140$ \\
                \textbf{Pregnant} & $743$\\
            \bottomrule
        \end{tabular}
    }	
	\label{tab:statistics}
    \vspace*{-0.6cm}
\end{table}

\vspace*{-0.3cm}
\section{Experimental Setup}
\label{sec:setup}

\myparagraph{Medical KG}
The MKG used in this work is based on \cite{liao2022medical},
which was constructed from a large number of Chinese drug instructions
containing detailed drug information such as treatments, ingredients, usage, drug-drug interactions, contraindications, or adverse reactions. 
%
It includes $113{,}094$ unique drugs ($47{,}367$ having DDIs), $30{,}026$ diseases, and $111{,}024$ ingredients. 



\vspaceopt
\myparagraph{Metrics}
We use \textit{Jaccard coefficient} (\textbf{Jaccard}),
\textit{precision} (\textbf{P}),
\textit{recall} (\textbf{R}),
\textit{F1-score} (\textbf{F1}), and
{\textit{DDI rate} (\textbf{DDI}).
In our settings, the \textit{DDI rate} is computed among the recommendations and concomitant drugs.}
All metrics are based on top-$5$ predictions and averaged.
Statistical significance against the best baseline (*) is measured using the paired
$t$-test with $p$<$0.05$.



\vspaceopt
\myparagraph{Baselines}
We compare \explainmr against a suite of baselines on our new \benchmark benchmark:

\squishlist
    \item \textbf{LLMs with ICL}.
        We compare against
        \textbf{GPT-4}~\cite{achiam2023gpt}, 
        \textbf{GLM-4-Plus}~\cite{glm2024chatglm},
        and \textbf{Qwen-Max}~\cite{yang2024qwen2}.
        We provide an instruction and select $8$ demonstrations for in-context learning (\textbf{+ ICL}).
        Demonstrations include queries and drug recommendations from the train set.
        The query is
        derived using patient information. 




    \item \textbf{DR methods}.
    We compare against {\textbf{GAMENET}~\cite{shang2019gamenet}, \textbf{LEAP}~\cite{zhang2017leap}, and \textbf{RETAIN}~\cite{choi2016retain}.
    We also adapt \textbf{DDxT}~\cite{alam2023ddxt} to our task for comparison,
    which was
    designed for predicting pathologies.}

    \item \textbf{Retrieval methods}. We include two retrieval baselines: \textbf{BM25} (top-$5$ drugs from Sec.~\ref{subsec:retrieval}) and \textbf{Semantic Similarity}~\cite{reimers-2019-sentence-bert}\footnote{\textcolor{urlcolor}{\url{https://huggingface.co/sentence-transformers/distiluse-base-multilingual-cased-v2}}}.

    \item \textbf{GNNs}. We construct a relational graph from the MKG,
    including
    details
    on the relevant drugs (Sec.~\ref{subsec:retrieval}),
    and apply two
    state-of-the-art
    GNNs:
    \textbf{R-GCN}~\cite{schlichtkrull2018modeling} and \textbf{GAT}~\cite{velivckovic2017graph}.

        
  \squishend      


\vspaceopt
\myparagraph{Configuration} 
The GNNs are trained with a batch size of $1$ for $5$ epochs (epoch-wise evaluation), learning rate of 
$10^{-5}$,
$500$ warm-up steps, weight decay of $0.01$, and AdamW optimizer.
We used $3$ GNN layers.
For multi-task learning, the weight for entity node classification was $0.7$ and $0.3$ for evidence nodes.
Overall, our GNN has $114$ million parameters, including the
LM encoder.

\begin{table}  
    \caption{
    Main results
    on the \textit{test} set of \benchmark.} 
    \vspace*{-0.4cm}
    \newcolumntype{G}{>{\columncolor [gray] {0.90}}c}
    \newcolumntype{H}{>{\setbox0=\hbox\bgroup}c<{\egroup}@{}}
    \newcolumntype{L}{>{}c}
	\resizebox*{0.90\columnwidth}{!}
    {
        \setlength{\tabcolsep}{3pt}
        \renewcommand{\arraystretch}{0.8}
        \footnotesize
        
        \begin{tabular}{l c c c c H c H}
    		\toprule
                \textbf{Method} 
                & \textbf{Jaccard} 
                & \textbf{P} & \textbf{R} & \textbf{F1} & 
               \textbf{Hit@1} & 
                \textbf{DDI} & \textbf{PRAUC}
                 \\
    	\midrule	    
 	
        \textbf{GPT-4 + ICL}
                &	$0.010$ 
                
                &	$0.041$ &	$0.010$ &	$0.013$  & $0.047$&$0.039$&$0.047$
                \\ 
                
                
            \textbf{GLM-4-Plus + ICL}
                &	$0.016$ 
                
                &	$0.057$ &	$0.018$ &	$0.020$  & $0.063$&$0.041$ &$0.066$
                \\ 

                \textbf{Qwen-Max + ICL}
                &	$0.020$ 
                 
                &	$0.062$ &	$0.027$ &	$0.028$ &$0.063$&$0.038$&$0.083$
                \\
                
                 
                 
                 
 	
            \midrule
             \textbf{DDxT}
                &$0.180$ 
                
                &	$0.306$ &	$0.209$ &	$0.229$ &$0.181$&$0.020$ &$0.318$
               \\
                
                
               \textbf{GAMENET}
                &$0.002$ 
                
                &	$0.010$ &	$0.004$ &	$0.001$ &$0.016$&$\boldsymbol{0.001}$ &$0.025$
               \\
               \textbf{LEAP}
                &$0.011$ 
                
                &	$0.011$ &	$0.018$ &	$0.009$ &$0.019$&$\boldsymbol{0.001}$ &$0.023$
               \\
               \textbf{RETAIN}
                &$0.013$ 
                
                &	$0.157$ &	$0.015$ &	$0.022$ &$0.182$&$\boldsymbol{0.001}$ &$0.198$
               \\

            \midrule
               \textbf{BM25} 
                &	$0.194$ 
                &	$0.363$ &	$0.469$ &	$0.286$ &$0.570$ &$0.043$ &$0.648$ \\
               \textbf{Semantic Similarity} 
                &	$0.216$ 
                &	$0.385$ &	$0.511$ &	$0.314$ &$0.647$ &$0.044$  &$0.697$\\
                 \midrule

                \textbf{R-GCN}
                &	{$0.195$} 
                	 
                &$0.381$ &	$0.477$ &	$0.291$ &$0.573$&$0.032$ &$0.662$
                \\
            \textbf{GAT}
                &	$0.208$ 
                &	$0.414$ &	$0.495$ &	$0.307$ &$0.642$ &$0.035$ &$0.713$
               \\


               \midrule

                	 
                
                
                
                \textbf{\explainmr} (ours)
                &	$\boldsymbol{0.253}$*  
                
                & $\boldsymbol{0.484}$*   &	$\boldsymbol{0.572}$* &	$\boldsymbol{0.361}$*  &$\boldsymbol{0.861}$ &$0.028$ &$\boldsymbol{0.877}$
                \\

    		\bottomrule
    	\end{tabular} 
    }

    \label{tab:main-res}
    \vspace*{-0.2cm}
\end{table}

\section{Experimental Results}
\label{sec:main-res}
\myparagraph{\method outperforms all baselines across most metrics}
Main results are shown in Table~\ref{tab:main-res}. 
First of all, we note that 
graph-based approaches generally
outperform LLMs and DR methods.
The reason is that GNNs better capture the
interactions
among entities. 
Second, the performance of retrieval methods is comparable to graph-based approaches; however, the DDI rate is higher, suggesting that while these methods effectively leverage patient information and drug similarities, but overlook potential DDIs.  
Third, 
\explainmr outperforms all baselines
across most
metrics
substantially.
Notably, \explainmr achieves an F1 score of $0.361$ higher than all other methods (second highest is $0.307$) and a low DDI rate $0.028$.
Interestingly, thanks to our
novel patient-attention mechanism,
\explainmr 
exhibits clear performance improvements compared to
popular
GNN
algorithms.

\vspaceopt
\myparagraph{DR methods struggle on \benchmark}
Though the methods achieve the lowest DDI rate, the performance on other metrics is poor, with F1 scores below $0.3$, suggesting these methods struggle on 
dealing with rare diseases or less frequently prescribed
drugs.



\vspaceopt
\myparagraph{LLMs fail with drug recommendation}
Despite the widespread application of LLMs in almost any domain,
we find that
all of the LLMs consistently fail on our challenging benchmark.
Less than {$0.03$} at F1 scores of the LLM predictions 
shows that LLMs
are far from an application-ready state for DR.

\begin{table} 
    \caption{Ablation studies on the \benchmark \textit{dev} set.} 
    \vspace*{-0.4cm}
    \newcolumntype{G}{>{\columncolor [gray] {0.90}}c}
    \newcolumntype{L}{>{}c}
    \newcolumntype{H}{>{\setbox0=\hbox\bgroup}c<{\egroup}@{}}
    \resizebox*{0.90\columnwidth}{!}
    {
        \setlength{\tabcolsep}{3pt}
        \renewcommand{\arraystretch}{0.8}
        \footnotesize
        
    	\begin{tabular}{l c c c c H c H}
    		\toprule
                \textbf{Method} 
               & \textbf{Jaccard} 
               & \textbf{P} & \textbf{R} & \textbf{F1} &
               \textbf{Hit@1}&\textbf{DDI}&\textbf{PRAUC}
                \\ 
            \midrule       
            \textbf{\explainmr}
                & {$\boldsymbol{0.252}$}	
               
               &	{$\boldsymbol{0.485}$} & {$\boldsymbol{0.568}$}  & {$\boldsymbol{0.359}$}  & {$\boldsymbol{0.850}$}&$\boldsymbol{0.028}$ &$\boldsymbol{0.866}$ \\
            \midrule
               

               \textbf{w/o patient-attention} 
                 &	{$0.210$} 
                 &	{$0.413$} & {$0.490$} & {$0.306$} &{$0.628$}&$0.018$&$0.703$\\
               

                 \textbf{w/o GNN} 
                 &	{$0.175$} 
                 &	{$0.362$} & {$0.406$} & {$0.258$} &{$0.636$}&$0.030$&$0.687$\\

    		\bottomrule
    	\end{tabular} 
    }
    \label{tab:ablations}
    \vspace*{-0.4cm}
\end{table}

\myparagraph{Ablation studies}
Ablation studies are shown in Table~\ref{tab:ablations}.
We removed
(i) our novel patient-attention, or (ii) replaced the GNN
by a cross-encoder~\cite{nogueira2019passage}.
Results verify that the
GNN
is an essential ingredient of \explainmr,
and that
especially our patient-attention
contributes substantially to the performance.


\vspaceopt
\myparagraph{Error analysis}
{\explainmr may
(i) 
miss relevant drugs 
in the top-$50$ BM25 results ($30.6\%$ of cases),
(ii) miss relevant drugs within its top-$5$ results  ($42.8\%$ of cases).
Future work could look into enhancing recall in the initial retrieval step or improve the fine-grained
ranking
potentially including information beyond MKGs.}

\vspaceopt
\myparagraph{Anecdotal example}
Table~\ref{tab:anecdote} shows an example of \method predicated recommendation with its evidence and a candidate identified
as contraindicated (as the patient is pregnant, highlighted in red), hence ranking relatively low (out of top-5).

\begin{table} [t] \footnotesize
	\centering
	\caption{Anecdotal example of \explainmr.}
    \vspace*{-0.4cm}
        \begin{tabular}{p{0.95\columnwidth}}
            \toprule
                \textbf{Patient}: 26 years old, Female, \textcolor{red}{\phrase{Pregnant}},
                Disease: \textcolor{blue}{\phrase{Soft tissue rheumatism}}, Symptoms: \phrase{Joint pain, Muscle aches}\\
            \midrule
                \textbf{Recommendation}: \phrase{Shogaol Soft Capsules}\\
                \textbf{Evidence}: Treatments: \phrase{\textcolor{blue}{Acute and chronic rheumatoid arthritis}}, Contraindications: \phrase{Use with caution in patients with diarrhea}, Ingredients: \phrase{Shogaol}\\
            \midrule
                \textbf{Candidate}: \phrase{Shennanxing Oral Liquid}\\
                \textbf{Evidence}: Treatments: \phrase{\textcolor{blue}{Muscle and joint aches and pains}}, Contraindications: \textcolor{red}{\phrase{Gestational period}}, \phrase{Allergic to ginseng}, Ingredients:{\phrase{Ginseng, Bupleurum}}\\
            \bottomrule
            
        \end{tabular}
	\label{tab:anecdote}
    \vspace*{-0.5cm}
\end{table}

\vspace*{-0.1cm}
\section{Related Work}
\label{sec:relatedwork}
Methods for DR can be roughly classified into two groups: longitudinal
and instance-based approaches. 
\textit{Longitudinal methods} exploit
patients’ longitudinal medical history. For example, ~\cite{choi2016retain, le2018dual} model EHRs as temporal sequences and apply RNNs to learn embeddings for drug predictions.
\cite{shang2019gamenet} uses the EHRs of a patient as the query and combines a DDI KG with a memory module implemented as a GCN for 
safe recommendations.
~\cite{tan20224sdrug} predicts drugs based on a diagnosis set similarity measurement model. These methods heavily rely on patients’ medical histories, falling short in cold start scenarios~\cite{wei2021contrastive} (i.e., new drugs/patients). 
\textit{Instance-based approaches} focus on the patient’s
current health status. For instance, \cite{zhang2017leap} utilizes attention
mechanisms to capture dependencies between drug labels in the current visit. 
\cite{zhao2024leave} leverages a pretrain-finetune learning
paradigm to enhance accuracy for rare diseases. 
\cite{kuang2024drugdoctor} trains a transformer-based encoder to capture the
relationship between drugs’ substructures and the disease of the current visit.
These methods achieve accurate recommendations for a single-visit scenario, but no traceable evidence to support the recommendations.


\vspace*{-0.1cm}
\section{Conclusion}
\label{sec:conclusion}
We proposed \explainmr, a novel DR approach operating over a large-scale medical KG.
\method
 enables drug recommendation for single-visit patients with rare diseases and \textit{tracing}
back these
recommendations
to the supporting
evidence.
Experiments
show the superiority of \explainmr compared to a wide range of competitive baselines.
We also release \benchmark, a new benchmark with a diverse set of patients and diseases,
for future work on DR.

\vspace*{0.1cm}
\myparagraph{Acknowledgements}
We thank Gerhard Weikum from the Max Planck Institute for Informatics for helpful feedback on the project.
The work was supported by the Major Science and Technology Projects in Sichuan Province (No.2024ZDZX0012).

\balance

\bibliographystyle{ACM-Reference-Format}
\bibliography{mr_clean}

\end{document}